\documentstyle[12pt,aasms4]{article}
\input epsf
\begin{document}
\title{On the equipartition of thermal and non-thermal
energy in clusters of galaxies}
\author{Pasquale Blasi}
\affil{Department of Astronomy and Astrophysics, The
University of Chicago,\\ 5640 S. Ellis Av., Chicago, IL60637}
\begin{abstract}
Clusters of galaxies are revealing themselves as powerful
sources of non thermal radiation in a wide range of wavelengths.
In order to account for these multifrequency observations
equipartition of cosmic rays (CRs) with the thermal gas in clusters
of galaxies is often invoked. This condition might suggest 
a dynamical role played by
cosmic rays in the virialization of these large scale structures
and is now testable through gamma ray observations.  We show here,
in the specific case of the Coma and Virgo clusters, for which 
upper limits on the gamma ray emission exist, that
equipartition implies gamma ray fluxes that are close or even in
excess of the EGRET limit, depending on the adopted model of CR injection. 
We use this bound to limit the validity of the equipartition
condition. We also show that, contrary to what claimed in 
previous calculations, the equipartition assumption implies gamma 
ray fluxes in the TeV range which can be detectable even by 
currently operating gamma ray observatories if the injection cosmic
ray spectrum is flatter than $E^{-2.4}$. 
\end{abstract}


\section{Introduction}

Several non-thermal processes have recently been detected in 
clusters of galaxies from the extreme ultraviolet (EUV) radiation in 
excess of the thermal expectation to the soft X-rays detected by 
ROSAT and BeppoSAX and again to hard X-ray excesses and radio
radiation. For the Coma cluster, by far the best studied cluster, 
a complete investigation of
the soft excess can be found in 
(Lieu et al. 1996a, 1996b; Bowyer, Lampton \&
Lieu 1996; Fabian 1996; Mittaz, Lieu \& Lockman 1998, Sarazin \&
Lieu 1998) 
\markcite{lieu1}\markcite{lieu2}
\markcite{bll96}\markcite{fab96}\markcite{mll98}\markcite{sarl98}
while the detection of the hard excess above 20 keV is reported in
(Fusco-Femiano et al. 1998)\markcite{fusco}. A recent review of the 
diffuse radio emission can be found in 
(Feretti et al. 1998) \markcite{feretti}. 
Some clusters show only emission in some region of 
frequency and not in others. Also for this reason Coma gives
the best possibility to make multiwavelengths studies.
A review of the current status of the multifrequency 
observations of Coma and viable models for the non thermal radiation
can be found in (Ensslin et al. 1998) \markcite{ensslin98}. As stressed 
by Fusco-Femiano et al. (1998)\markcite{fusco}, if the hard X-ray excess 
is due to inverse compton scattering (ICS) off the photons of the
microwave background, then the combined radio and 
hard X-ray observations of Coma 
imply a small value for the average intracluster magnetic field, of
order $B\simeq 0.1-0.2\mu G$. Such a small value of the field requires
large energy densities in electrons, and, as pointed out
in (Lieu et al. 1999)\markcite{lieu}, CR energy densities comparable  
with the equipartition value are required. This conclusion is only weakly 
dependent on
the specific model (primary or secondary) for the production of 
the electrons responsable for the radiation. In fact the need for 
large CR energy densities
was recently confirmed by Blasi \& Colafrancesco (1999)\markcite{blasi}, in 
the context of the 
secondary electron model. Lieu et al. (1999)\markcite{lieu} also correctly 
pointed out that the assumption of equipartition is limited by the
production of gamma rays through neutral 
pion decay, but this flux was claimed to be much smaller than the 
EGRET sensitivity, falling below the EGRET upper 
limit already imposed on the gamma ray flux from the Coma and Virgo clusters
(Sreekumar et al. 1996)\markcite{sreek}.

We calculate here the flux of gamma rays from the Coma 
and Virgo clusters in the
assumption of equipartition of CRs with the thermal energy in the
cluster, for two different models of the CR injection in the 
intracluster medium (ICM), and for different injection spectra
and find that in some cases the gamma ray flux is in excess
of the EGRET limit. Moreover, we find that for injection CR spectra flatter
than $E^{-2.4}$ (for $E\gg m_p c^2$) some currently operating experiments
like STACEE, HEGRA and Whipple could detect the gamma ray signal from Coma
and Virgo in the TeV range, provided the CRs are in equipartition, or 
put strong
constraints on this condition if no signal is detected.

The paper is planned as follows: in section \ref{gamma} we outline
the calculations of the gamma ray fluxes from clusters; in section
\ref{cr} we describe the models of CR propagation that we used and in
section \ref{results} we describe our results for the Coma and Virgo clusters.

\section{The gamma ray fluxes}\label{gamma}

In this section we calculate the flux of gamma rays due to the
decay of neutral pions produced in CR collisions in the ICM. This
channel provides the dominant contribution to gamma rays above
$100$ MeV. 

Independent of the sources that provide the CRs in clusters, the
equilibrium CR distribution is some function $n_p(E_p,r)$ of the
proton energy $E_p$ and of the position in the cluster. For
simplicity we assume the cluster to be spherically symmetric,
so that the distance $r$ from the center is the only space
coordinate. We determine $n_p$ for different injection models
in the next section.
The rate of production of gamma rays with energy $E_\gamma$
per unit volume at distance $r$ from the cluster center is
given by Blasi \& Colafrancesco (1999) \markcite{blasi}
\begin{equation}
q_\gamma(E_\gamma,r)=2 n_H(r) c
\int_{E_\pi^{min}(E_\gamma)}^{E_p^{max}}
dE_\pi \int_{E_{th}(E_\pi)}^{E_p^{max}}
dE_p F_{\pi^0}(E_\pi,E_p)\frac{n_p(E_p,r)}{(E_\pi^2+m_\pi^2)^{1/2}},
\label{eq:gamma1}
\end{equation}
where $E_\pi$ is the pion energy, $E_\pi^{min}=E_\gamma+
m_\pi^2/(4E_\gamma)$ is the minimum pion energy needed to 
generate a gamma ray photon with energy $E_\gamma$ and $E_p^{max}$
is some maximum energy in the injected CR spectrum (our 
calculations do not depend on the value of $E_p^{max}$).
Here $n_H(r)$ is the density of thermal gas at distance $r$
from the cluster center. For Coma we model the gas density 
through a King profile:
\begin{equation}
n_H(r)=n_0 \left[1+\left(\frac{r}{r_0}\right)^2\right]^{-3\beta/2},
\label{eq:den_coma}
\end{equation}
where $r_0\approx 400$ kpc is the size of the cluster core, 
$n_0\approx 3\times 10^{-3}~cm^{-3}$  
and $\beta$ is a phenomenological parameter in the
range $0.7-1.1$ (Sarazin 1988)\markcite{sarazin} (we use $\beta=0.75$).\par
For Virgo, we fit the gas density profile given by
Nulsen \& Bohringer (1995)\markcite{nb95} to find 
\begin{equation}
n_H(r)=0.076\cdot \left(\frac{r}{4.8kpc}\right)^{-1.16}~cm^{-3}.
\label{eq:den_virgo}
\end{equation}
The function $F_\pi$ in eq. (\ref{eq:gamma1}) represents the 
cross section for the production of neutral pions
with energy $E_\pi$ in a CR collision at energy $E_p$
in the laboratory frame. Determining this function is
complicate in the low energy regime where 
data is scarse. 
A possible approach was proposed by Dermer (1986)\markcite{dermer}
and recently reviewed by Moskalenko \& Strong (1998)  
\markcite{strong} and is based on the
isobar model. This approach is valid for CR collisions at
$E_p\leq 3$ GeV and consists in treating the pion production
as a process mediated by the generation and decay of the resonance
$\Delta(1232)$ in the $pp$ interaction. We refer to the papers
by Dermer (1986) and Moskalenko \& Strong (1998) for the 
detailed expressions for $F_\pi$.
For $E_p\geq 7$ GeV the scaling approach is
an excellent approximation of the function $F_\pi$. In this
regime the differential cross section for $pp$ collisions can
be written as
\begin{equation}
\frac{d\sigma}{dE_\pi}(E_p,E_\pi)=\frac{1}{E_\pi}\sigma_0
f_\pi(x)
\label{eq:scaling}
\end{equation}
where $x=E_\pi/E_p$, $\sigma_0=3.2\times 10^{-26}~cm^2$ 
and $f_\pi(x)=0.67(1-x)^{3.5}+0.5e^{-18x}$ 
is the so called scaling function. In the scaling regime, 
the function $F_\pi$ coincides with the differential cross 
section given in eq. (\ref{eq:scaling}). \par
Once the gamma ray emissivity is known from eq. (\ref{eq:gamma1}),
the flux of gamma rays with energy $E_\gamma$ is simply given by
volume integration
\begin{equation}
I_{\gamma}(E_\gamma)=\frac{1}{4\pi d^2}\int_0^{R_{cl}} dr 4\pi r^2
q_\gamma(E_\gamma,r)
\label{eq:flux}
\end{equation}
where $d$ is the distance to the cluster and $R_{cl}$ is the cluster
radius. In fact $R_{cl}$ here plays the role of the size of the region
where the non thermal processes are observed. We adopt here the value
suggested from radio observations in Coma, $R_{cl}\approx 1$ Mpc.
This is however a very conservative case and it seems likely that 
magnetic fields extend to larger regions. In fact in Ensslin,
Wang, Nath \& Biermann (1998a) \markcite{ensslin98bis}
the injection of energy due to formation 
of black holes in the Coma cluster was estimated and compared with the
thermal energy in a region of $5h_{50}^{-1}$ Mpc ($h_{50}=h/0.5$). 
For the Coma cluster
we shall also consider this less conservative case.

\section{The Cosmic Ray Distribution}\label{cr}

Several sources of CRs in clusters of galaxies were discussed by
Berezinsky, Blasi \& Ptuskin (1997)
\markcite{bbp} and it was argued that the known sources (AGNs, radiogalaxies,
accretion shocks) are not able to provide CRs in equipartition with the
thermal gas. An intense and short period of powerful emission from
the cluster sources was also considered, consistently with the observed
iron abundance in the cluster, with the same conclusion.
Since recent observations of non thermal radiation from
clusters seem to suggest that equipartition is indeed required, we 
do not make here any assumption on the type of sources and instead 
we assume equipartition and analyze the observational consequences
of this assumption.\par
The equipartition energy can be easily estimated from the total 
thermal energy of the gas, assuming it has a temperature $T$:
\begin{equation}
E_{eq}\approx\frac{3}{2}k_B T\int_0^{R_{cl}} dr 4\pi r^2 n_H(r) 
\label{eq:equi}
\end{equation}
where $n_H(r)$ is given by eq. (\ref{eq:den_coma}) for the
Coma cluster and by eq. (\ref{eq:den_virgo}) for the Virgo cluster. 
The temperature adopted for Coma is $T_{Coma}=8.21$ k while for 
Virgo we used $T_{Virgo}=1.8$ k (Nulsen \& Bohringer 1995)\markcite{nb95}.
Therefore, from the previous equation we obtain 
$E_{eq}=1.6\times 10^{63}$ erg for Coma and $E_{eq}=1.5\times 10^{62}$
erg for Virgo. These numbers could underestimate the total 
thermal energy due to the contribution of gas out of the $\sim 1$ Mpc
region. In fact Ensslin et al. (1998a) \markcite{ensslin98bis} 
estimated for the Coma cluster
that the thermal energy in a region of $5h_{50}^{-1}$ Mpc is $\sim 1.3\times
10^{64}h_{50}^{-5/2}$ erg, a factor $\sim 6$ larger than estimated above.
They also calculate the expected injection of total (thermal plus
non thermal) energy due to black hole formation in the cluster,
and find in the same region a similar number. 
  
Since not only the energy budget in CRs is not known, but 
also their spatial distribution is very poorly constrained, 
we consider here two extreme scenarios for the injection of CRs and we 
calculate the equilibrium CR distribution from the transport equation.

{\it i) Point source}\par
As argued by Berezinsky, Blasi \& Ptuskin (1997)\markcite{bbp},
Colafrancesco \& Blasi (1998)\markcite{cb} and Ensslin et al. (1998)
\markcite{ensslin98} it is likely that for most of the
cluster's age the main contributors to CRs in clusters are located in the 
cluster core. This is the case if a radiogalaxy or more generally a 
powerful active galaxy or a shock produced by merging is the source/
accelerator of CRs. There is an additional argument that plays in favor 
of a source mainly concentrated in the center of the cluster: if the
average spatial distribution of the galaxies in a cluster is not a 
strong function of time, then it is reasonable to assume that at all times,
as today, the distribution of the sources is peaked around the
cluster center. According with Ensslin et al. (1998a) \markcite{ensslin98bis}
(see also references
therein) the spatial distribution of galaxies in Coma is well represented by 
a King-like profile $n_{gal}(r)=[1+(r/r_g)^2]^{-0.8}$, with 
$r_g\simeq 160$ kpc, appreciably smaller that the cluster core, so that a 
source concentrated in the center seems a reasonable assumption.

Therefore we assume that the source can be modelled 
as a point source with a rate of injection of CRs given by a power
law in momentum $Q(E_p)=Q_0 p_p^{-\gamma}$, where 
$p_p=\sqrt{E_p^2-m_p^2}$ is the CR momentum and the normalization constant
is determined by energy integration 
\begin{equation}
Q_0\int_0^{E_p^{max}} dT_p T_p p_p^{-\gamma} = L_p,
\end{equation}
where $T_p$ is the kinetic energy and $L_p$ is the CR luminosity at injection,
forced here to be correspondent to the establishment of equipartition
in the cluster. We estimate it averaging the equipartition energy on the
age of the cluster: $L_p\approx E_{eq}/t_0$.\par
The transport equation that gives the distribution of CRs at distance $r$
from the source and after a time $t$, namely $n_p(E_p,r,t)$, can 
be written in the form
\begin{equation}
\frac{\partial n_p(E_p,r,t)}{\partial t} - D(E_p) \nabla^2 n_p(E_p,r,t) -
\frac{\partial}{\partial E_p}
\left[ b(E_p) n_p(E_p,r,t)\right] = Q(E_p)\delta (\vec {r}),
\label{eq:transport}
\end{equation}
where $D(E_p)$ is the diffusion coefficient and $b(E_p)$ is
the rate of energy losses.\par
As shown in (Berezinsky, Blasi \& Ptuskin 1997\markcite{bbp}, 
Colafrancesco \& Blasi 1998 \markcite{cb}, Blasi \& Colafrancesco 1999\markcite{blasi})
for CR protons the energy losses can
be neglected and eq. (\ref{eq:transport}) has the simple solution
(Blasi \& Colafrancesco 1999) \markcite{blasi}
\begin{equation}
n_p(E_p,r,t)=\frac{Q_p(E_p)}{D(E_p)}\frac{1}{2\pi^{3/2} r}
\int_{r/r_{max(E_p)}}^{\infty} dy e^{-y^2}.
\label{eq:sol}
\end{equation}
where $r_{max}(E_p)=\left[4D(E_p)t\right]^{1/2}$ is the maximum 
distance that on average particles with energy $E_p$ could diffuse away
from the source in the time $t$. We are interested here in the case $t=t_0$
($t_0$ here is the age of the cluster, taken as comparable with the 
age of the universe).
The solution of the equation $r_{max}(E_p)=R_{cl}$ gives an estimate
of the maximum energy $E_{max}$ for which CRs can be considered confined in 
the cluster volume for all the age of the cluster. 
For reasonable choices of the diffusion coefficient the confined CRs provide
the main contribution to the energy budget of CRs in clusters and also
to the integral flux of gamma rays above $100$ MeV, calculated as 
explained in the previous section. As far as gamma rays produced by 
interactions of confined CRs are concerned
the flux of gamma radiation is independent on the choice of the 
diffusion coefficient, as pointed out by Berezinsky, Blasi \&
Ptuskin (1997)\markcite{bbp}, and the spectrum 
of gamma rays simply reflects the spectrum of the parent protons
(for $E_\gamma\geq 1$ GeV).
Rigorously this is true only for spatially constant 
intracluster gas 
density, while a density profile, as assumed here, results in a 
weak dependence of the gamma ray spectrum on the diffusion details.
Therefore, for the sake of completeness 
we adopt here a specific choice of the diffusion coefficient:
we assume that the fluctuations in the magnetic field in the cluster
are well represented by a Kolmogorov power spectrum $P(k)\propto k^{-{5/3}}$
and we calculate the diffusion coefficient according with the
procedure outlined by Colafrancesco \& Blasi (1998)\markcite{cb}, which gives
\begin{equation}
D(E_p)=2.3\times 10^{29} E_p(GeV)^{1/3} B_\mu^{-1/3} 
\left(\frac{l_c}{20kpc}\right)^{2/3} ~ cm^2/s
\label{eq:diff}
\end{equation}
where $B_\mu$ is the value of the magnetic field in $\mu G$ and
$l_c$ is the scale of the
largest eddy in the power spectrum of the magnetic field.\par
Eqs. (\ref{eq:diff}) and (\ref{eq:sol}) completely define the distribution
of cosmic rays in the cluster in the case of a point source.

{\it i) Spatially homogeneous injection}\par

As pointed out above, the budget of CRs in clusters is largely dominated
by confined CRs, so that in the case of spatially homogeneous injection 
the distribution of CRs can be easily written in the form
\begin{equation}
n_p(E_p,r)=n_0 \frac{\epsilon_{tot}}{V} p_p^{-\gamma}
\label{eq:homo}
\end{equation}
where $V=(4/3)\pi R_{cl}^3$ is the injection volume, $\epsilon_{tot}$ 
is the total energy injected in the cluster in the form of CRs and $n_0$ is 
calculated by the normalization condition
\begin{equation}
n_0\int_0^{E_p^{max}} dT_p T_p p_p^{-\gamma} = E_{eq},
\end{equation}
where $E_{eq}$ is calculated according to eq. (\ref{eq:equi}). Clearly
eq. (\ref{eq:homo}) does not describe well the CR distribution very close to
the cluster boundary. Moreover at sufficiently high energy, where CRs
are not confined in the cluster volume the CR spectrum suffers a 
steepening to $E_p^{-(\gamma+\eta)}$, with $\eta=1/3$ for a Kolmogorov
spectrum. 

\section{Results and conclusions}\label{results}

We study the observational consequences of the assumption of equipartition 
between CRs and thermal gas in clusters of galaxies. In particular
we calculated the flux of gamma radiation from the Coma and 
Virgo clusters when 
equipartition is assumed. 
This assumption seems to be required if a ICS 
origin is accepted for the hard and soft X-ray excess and for
the EUV flux from Coma and other clusters of galaxies (note however 
that alternative possibilities can be proposed). 
In particular, according to Lieu et al. (1999)\markcite{lieu}, 
{\it in order to account for the observed cluster
soft excess flux from Coma, equipartition between CRs and gas 
is unavoidable}. In (Berezinsky et al. (1997))\markcite{bbp} different 
possible models of CR 
injection in clusters were considered, including active galaxies,
accretion shocks during the formation of the cluster and a possible
bright phase in the past of the cluster galaxies, but none of these
sources could account for CR energy densities larger that $1-5\%$ of the
equipartition value, if a conversion efficiency of $\sim 10\%$ was 
assumed for the injection of non thermal energy from the
total energy of the sources.

On the other hand Ensslin et al. (1998a) \markcite{ensslin98bis}
compared the thermal energy 
in a $5 h_{50}^{-1}$ Mpc region with the energy injected during the 
formation of massive black holes in the Coma cluster. The total energy
(thermal plus non thermal) 
released in this process was estimated to be comparable with the thermal
energy in the cluster (if an efficiency factor is
assumed, the non thermal energy may be smaller than the equipartition
value).

Since our knowledge of the sources of CRs in 
clusters is still very poor, we decided to adopt here a phenomenological 
approach and try to find observational tests or consequences 
of our assumptions. The most striking consequence of a large abundance 
of CRs in a cluster is the production of gamma rays through the generation
and decay of neutral pions in $pp$ interactions. Since the EGRET 
instrument put an upper limit on the flux of gamma radiation above 
$100$ MeV from Coma and Virgo ($F_\gamma^{EGRET}(>100MeV)\lesssim
4\times 10^{-8}~phot/(cm^2 s)$ (Sreekumar et al. 1996\markcite{sreek})), 
we can use this constraint to test the equipartition assumption.\par
Our calculations were carried out for two extreme models of injection of
CRs in the cluster, namely a point source in the cluster core and a 
spatially homogeneous injection in the cluster volume. In the case of
a point source, we can think of it as an {\it effective source}, in the
sense that on average a dominant source or a set of sources are 
located at the cluster center. 
In this sense it is not needed that the same source remains active 
for all the age of the cluster.\par
The energy spectrum of the injected CRs was assumed to be a power
law in momentum, as expected for a shock acceleration
spectrum, and two extreme values of the power index were studied, namely
$\gamma=2.1$ and $\gamma=2.4$, which encompass the whole range of power
indexes expected from shock acceleration (other models of acceleration also
give power laws in the same range of parameters).\par
Since the gamma ray spectra depend (although very weakly) on the choice
of the diffusion coefficient, we made here a specific choice, modelling
the spectrum of fluctuations of the field by a Kolmogorov spectrum and
calculating the diffusion coefficient according with eq. (\ref{eq:diff}).
In the numerical calculations we used $B_\mu=0.1$ and $l_c=20$ kpc (if
for instance we use $B_\mu=1$ the results on the integral fluxes change 
only by $\sim 10\%$, confirming the weak dependence on the diffusion
details mainly due to the use of a specific gas density profile).
 
The integral fluxes of gamma radiation above $100$ MeV for the 
cases mentioned above and in the conservative scenario of $R_{cl}=1$ Mpc, 
are reported in Table \ref{table} for the Coma and Virgo clusters. 

\placetable{table}

Due to the appearence 
of the flat region at low gamma ray energy, typical of spectra from pion
decay, there is not a strong dependence of the integral flux on $\gamma$. 
In some cases considered the gamma ray flux exceeds the EGRET upper limit. 
As it could be expected, the gamma ray flux is larger for the case of a 
point source in the cluster center and the EGRET limit is exceeded by 
a factor $\sim 1.7$
for Coma and by a factor $\sim 9$ for Virgo. 
In the case of homogeneous injection the gamma ray fluxes are slightly
smaller than the EGRET upper limit both for Coma and Virgo. 

These results are more impressive when the condition of equipartition 
is imposed on a larger region of size $5h_{50}^{-1}$ Mpc (Ensslin et al.
(1998a)\markcite{ensslin98bis}): for a single source in the cluster
center the EGRET limit is exceeded by $\sim 9$ for $\gamma=2.1$ and
by $\sim 8$ for $\gamma=2.4$. For an homogeneous injection the predicted
fluxes are in excess of the EGRET limit by $\sim 7$ for $\gamma=2.1$ and 
by $\sim 6$ for $\gamma=2.4$.

It is worthwhile to stress again that this result is practically
independent on the specific choice of the diffusion coefficient. 
In fact, the CRs relevant
for the production of gamma rays in the energy range $0.1-10$ GeV
are certainly confined in the cluster for any reasonable choice of the 
diffusion coefficient, and the spectrum in this region is independent 
on this choice. The effects of the diffusion may appear only at higher
energy where gamma rays are produced by CRs not confined in the cluster. 
As shown in (Berezinsky et al. 1997)\markcite{bbp} this produces a 
steepening of the gamma ray spectrum
to a power law with index $\gamma+\eta$ (with $\eta=1/3$ in the case of
a Kolmogorov spectrum of fluctuations) at energies larger than a {\it knee}
energy $E_K$.
At smaller energies the gamma ray 
spectrum reproduces the spectrum of the parent CRs. The transition appears
at $E_K\propto \left[R_{cl}^2/(B_\mu^{-1/3} t_0)\right]^{1/\eta}$, 
as obtained from 
the equation $r_{max}(E_p)=R_{cl}$. Actually this was 
shown in (Berezinsky et al. 1997)\markcite{bbp} for the case of a 
constant intracluster gas density. In the more
realistic case considered here, where the gas is modelled by a King 
or a power law profile, the gamma ray spectrum suffers a smooth 
steepening even for confined CRs, but this affects the 
integral flux above $100$ MeV only at the level of $\sim 10\%$.\par
The integral spectra of gamma rays from Coma with energy 
$>E_\gamma$ as functions
of the energy $E_\gamma$ are shown in Fig. 1a (for the point source)
and 1b (for the homogeneous case) for $R_{cl}=1$ Mpc. 
In the same plot we draw the sensitivity
limits for several present and planned experiments for gamma ray astronomy.
The solid lines refer to $\gamma=2.1$ while the dashed lines are
obtained for $\gamma=2.4$.

\placefigure{fig1}

The fluxes in the 
energy region $E_\gamma<100$ GeV are well above the detectability limit
of GLAST, so that there is no doubt that the question of equipartition will 
be completely answered with the next generation gamma ray satellites. 
However fig. 1 also shows that the signal from 
Coma could be  
detectable even in some current experiments, provided $\gamma\leq 2.4$. In 
particular STACEE could detect the signal above $30$ GeV and Whipple might
detect the signal for $E_\gamma\geq 250$ GeV. The flux should be
detectable by the HEGRA Cerenkov telescope above $500$ GeV. A non-detection
from these experiments would imply a reduction of the energy density
in clusters by about one order of magnitude below equipartition for 
$\gamma=2.1$. For steep spectra only STACEE has a slim 
chance to detect 
the signal. In the same energy range the next generation gamma ray experiments
will very likely measure the flux of gamma rays for any value of $\gamma$
in the range considered here.\par
In the case of the Virgo cluster and $R_{cl}=1$ Mpc, the fluxes are 
plotted in fig. 2a (for sources in the center) and fig. 2b 
(for a homogeneous injection) and conclusions
similar to the ones outlined for Coma hold.

Note that this result is subtantially different
from the one obtained in previous calculations. In particular 
Ensslin et al. (1997)\markcite{ensslin97}
reached the conclusion that the fluxes from Coma and Virgo are {\it orders
of magnitude too low to be detectable} in the TeV range. This conclusion
was obtained because following
Dar and Shaviv (1995)\markcite{dar} the gamma ray spectrum 
was assumed
to reproduce the equilibrium spectrum of CRs in the Galaxy 
$\propto E_\gamma^{-2.7}$ (this did not affect appreciably their
integral fluxes above 100 MeV, which are not very different from
the ones obtained here for the homogeneous case).
However, as shown in (Berezinsky et al. 1997\markcite{bbp}, 
Colafrancesco \& Blasi 1998\markcite{cb}, Blasi \& Colafrancesco 1999
\markcite{blasi}) and confirmed here, the spectrum 
of gamma rays from $pp$ collisions in clusters does not reproduce the 
equilibrium CR spectrum , but the generation spectrum 
as far as gamma ray photons are produced by interactions of CRs
confined in the cluster, as it is the case for gamma rays with energy 
less than $\sim 1-10$ TeV, for the values of the parameters used here 
(though, as pointed 
out before, a slight steepening is introduced by the gas density profile). 
Therefore the gamma ray spectrum from CRs in clusters is approximately
$E_\gamma^{-\gamma}$ up to some maximum energy $E_K$ where CRs begin to be
no longer confined in the cluster volume. As a consequence the gamma ray
fluxes in the TeV range could be detectable even by present experiments
if the CRs are in equipartition with the gas in the cluster. On the other
hand, if no flux is detected, this will put a strong constraint 
on the equipartition assumption.\par

While the low energy integral gamma ray flux is very weakly 
dependent on the choice of the diffusion coefficient, the correspondent
flux at higher energies is more sensitive to it, since, as explained
above, the position of the knee is affected by this choice. In the
context of a Kolmogorov spectrum, the maximum diffusion coefficient is
obtained for a larger value of $l_c$. The choice $l_c\approx 20$ kpc was 
inspired by the typical size of the galaxies in the cluster. The largest
scale where the magnetic fluctuations are injected is the typical distance 
between galaxies, of the order of $l_c\approx 100$ kpc. For $l_c\approx
20$ kpc the position of the knee is at $\sim 1-10$ TeV, while for 
$l_c\approx 100$ kpc the knee is at $\sim 10-20$ GeV. However, since
the steepening in the gamma ray spectrum begins at large energy,
the difference in the plots caused by the use of this larger diffusion
coefficient is a factor $\sim 2$ at $\sim 100$ GeV (for $R_{cl}=1$ Mpc), 
so that the 
possibility of detecting the gamma ray fluxes in this energy region
is not appreciably affected and remains an interesting possibility.
The situation improves rapidly
with an increasing $R_{cl}$. In fact the value of the knee energy for
a Kolmogorov spectrum goes like $R_{cl}^6$ and high energy CRs  
are easily confined in a region of $4-5$ Mpc.
If, following Ensslin et al. (1998a)\markcite{ensslin98bis},
we use $R_{cl}=5h_{50}^{-1}$ Mpc, then, as shown before the 
absolute gamma ray fluxes increase at all energies by a factor
$\sim 5-10$ and the steepening at high energy is only found at 
$E\gtrsim 10^3$ TeV for the diffusion coefficient in eq. (\ref{eq:diff}).

At present we can only use the EGRET limit as a constraint. For Coma 
this limit implies that the energy density must be smaller than $\sim 60\%$ of
the equipartition value if the CRs are mainly contributed by sources in 
the central part of the cluster (with $R_{cl}=1$ Mpc). 
If injection occurs uniformly over
the cluster volume, than the equipartition CR energy density 
is compatible with the EGRET limit (for $R_{cl}=1$ Mpc).
For Virgo cluster the EGRET limit implies that 
the energy density in CRs must be smaller than 
$\sim 10\%$ of the equipartition
value for the case of sources in the center.

As stressed above, the case of equipartition in a larger region, 
(Ensslin et al. (1998a)) is already ruled out 
by present gamma ray observations by EGRET. As a consequence the CR 
energy density in this case for Coma is forced to be $\lesssim 12\%$
of the equipartition value for a point source and $\lesssim 14\%$
for homogeneous injection.

\placefigure{fig2}

We suggest that experiments like STACEE, HEGRA, Whipple and future
gamma ray
experiments look at the signal from nearby clusters, because this could
definitely confirm or rule out the possibility that equipartition of 
CRs with the thermal gas is achieved in clusters of galaxies, or at
least impose new and stronger constraints on the maximum allowed 
CR energy density in clusters.\par

\acknowledgments
The author is grateful to A. Olinto, S. Colafrancesco, C. Covault and
R. Ong for many useful discussions and to the anonymous referee for 
several interesting comments. The research of P.B. is funded through 
a INFN fellowship at the University of Chicago.

\newpage
\begin{center}
\begin{table}
\caption{Summary of the gamma ray integral fluxes at $E_\gamma>100$ MeV
\label{table}}
\vskip 0.3truecm
\begin{tabular}{| c | c | c |}
\hline
\hline
$Model$ & $\Phi_{Coma}(E>100MeV)$ $phot/(cm^2 s)$& 
$\Phi_{Virgo}(E>100MeV)$ $phot/(cm^2 s)$ \\ \hline

$\gamma=2.1$, Point source & $7\times 10^{-8}$ & $3.7\times 10^{-7}$  \\
$\gamma=2.4$, Point source & $6.5\times 10^{-8}$ & $3.6\times 10^{-7}$  \\
$\gamma=2.1$, Homogeneous  & $2.5\times 10^{-8}$ & $3\times 10^{-8}$  \\
$\gamma=2.4$, Homogeneous  & $2.1\times 10^{-8}$ & $2.6\times 10^{-8}$  \\

\hline
\hline
\end{tabular}
\end{table}
\end{center}

\newpage
\figcaption[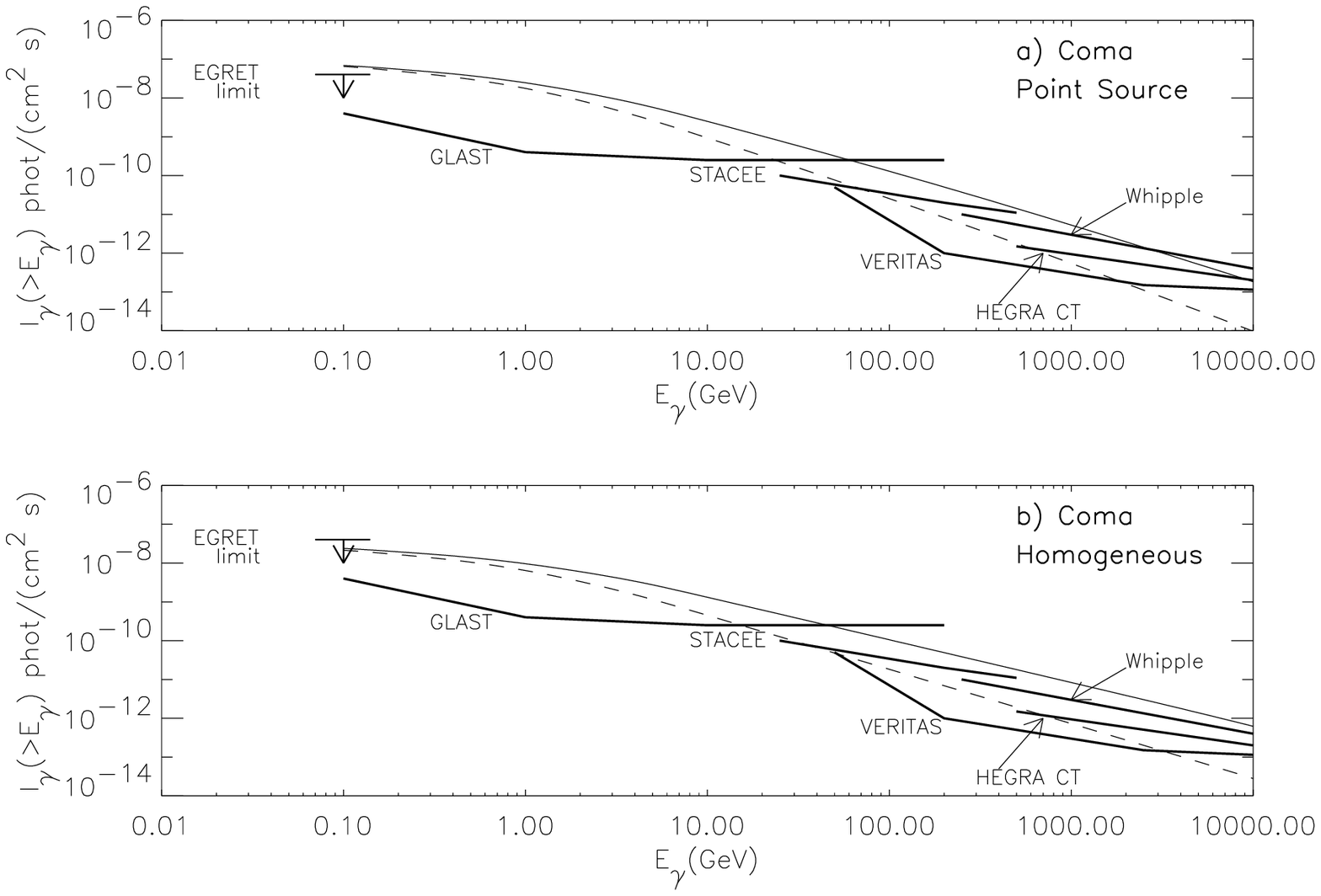]{Integral fluxes of gamma radiation
from the Coma cluster compared with the sensitivity limits
of some current and proposed gamma ray experiments (thick solid lines). 
The thin solid
line curves are referred to $\gamma=2.1$ and the dashed ones to
$\gamma=2.4$. In both cases $R_{cl}=1$ Mpc.
The EGRET limit at $100$ MeV is indicated by the
arrow. {\it a)} Injection of CRs by a point source in the
cluster center. {\it b)} Homogeneous injection of CRs over the
cluster volume.\label{fig1}}

\figcaption[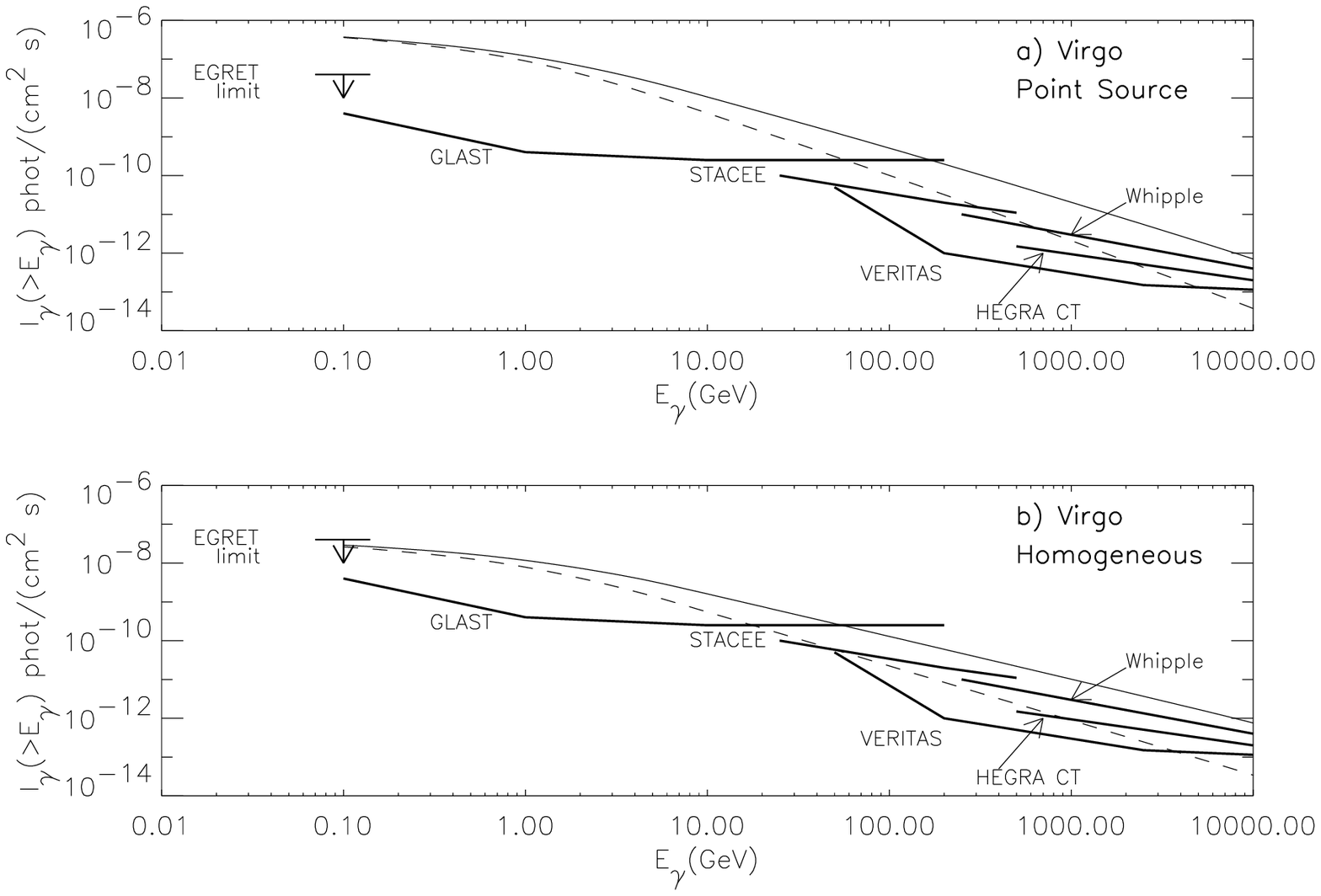]{The same as Fig. 1 for the case of 
the Virgo cluster. \label{fig2}}



\begin{references}


\reference{bbp}
Berezinsky, V.S., Blasi, P., \& Ptuskin, V. S. 1997, \apj, 487, 529.

\reference{blasi}
Blasi. P., \& Colafrancesco, S. 1999, to appear in Astrop. Phys.

\reference{bll96}
Bowyer, S., Lampton, M., \& Lieu, R. 1996, Science, 274, 1338.

\reference{cb}
Colafrancesco, S., \& Blasi, P. 1998, Astrop. Phys., 9, 227.

\reference{dar}
Dar, A., \& Shaviv, N. J. 1995, \prl, 75, 30, 52.

\reference{dermer}
Dermer, C. D. 1986, \aap, 157, 223.

\reference{ensslin97}
Ensslin, T. A., Biermann, P. L., Kronberg, P. P., \& Wu, X.-P. 1997,
\apj, 477, 560.

\reference{ensslin98}
Ensslin, T. A., Lieu, R., \& Biermann, P. L., preprint astro-ph/9808139.

\reference{ensslin98bis}
Ensslin, T., Wang, Y., Nath, B. B., \& Biermann, P. L. 1998a, \aap, 333, L47.

\reference{fab96}
Fabian, A. C. 1996, Science, 271, 1244.

\reference{feretti}
Feretti, L., \& Giovannini, G. 1997, Contribution to the Rencontres
Astrophysiques International Meeting {\it A New Vision of an Old
Cluster: Untangling Coma Berenices}, eds. F. Durret et al., held in Marseille,
France 17-20 June 1997 (astro-ph/9709294).

\reference{fusco}
Fusco-Femiano, R., et al. 1999, preprint astro-ph/9901018.

\reference{lieu1}
Lieu, R., Mittaz, J. P. D., Bowyer, S., Breen , J. O., Lockman, F. J.,
Murphy, E. M. \& Hwang, C.-Y. 1996a, Science, 274, 1335.

\reference{lieu2}
Lieu, R., Mittaz, J. P. D., Bowyer, S., Lockman, F. J.,
Hwang, C.-Y., \& Schmitt, J. H. M. M. 1996b, \apj, 458, L5.

\reference{lieu}
Lieu, R., Ip, W.-H., Axford, W. I., \& Bonamente, M. 1999, \apj, 510, L25.

\reference{mll98}
Mittaz, J. P. D., Lieu, R., \& Lockman, F. J. 1998, \apj, 498, L17.

\reference{strong}
Moskalenko, I. V. \& Strong, A. W 1998, \apj, 493, 694.

\reference{nb95}
Nulsen, P. E. J., \& Bohringer, H. 1995, \mnras, 274, 1093.

\reference{sarazin}
Sarazin, C. L. 1988, {\it X-ray emission from clusters of galaxies}, 
Cambridge University Press, Cambridge.

\reference{sarl98}
Sarazin, C. L., \& Lieu, R., Ip 1998, \apjl, 494, L177.

\reference{sreek}
Sreekumar, P. et al. 1996, \apj, 464, 628.


\end{references}
\end{document}